\documentclass[twocolumn,showpacs,amsmath,amssymb]{revtex4}

\usepackage{graphicx}

\begin{document}

\title{Applying the Dirac equation to derive the transfer matrix for
piecewise constant potentials}

\author{Ion I. Cot\u aescu \email{cota@physics.uvt.ro}}
\author{Paul Gravila \email{gravila@physics.uvt.ro}}
\author{Marius Paulescu \email{marius@physics.uvt.ro}}

\affiliation{West University of Timi\c soara, \\
       {\small  V. P\^ arvan Ave. 4, RO-300223 Timi\c soara}}

\date{\today}

\begin{abstract}

One propose a relativistic version of the transfer matrix method
for an electron moving through a given number of rectangular
barriers of arbitrary shape. It is shown that starting with the
Dirac equation depending on the effective mass and a suitably
chosen relativistic potential, one obtains a relativistic transfer
matrix which takes the correct traditional form in the
non-relativistic limit.
\end{abstract}


\maketitle

\section{Introduction}

The recent advances in nanostructure technology allow the
fabrication of electronic devices which inherently experience
quantum effects in their operation. There is a need for
theoretical tools, not only to understand the behavior of actual
nanoscale devices, but also to initiate future research and
developments.

The simplest non-relativistic quantum modeling of nanoscale
semiconductor devices is based on the Schr\"{o}dinger equation
written in solid state domains where the potential is constant and
the influence of the lattice is encapsulated in the value of
effective electron mass. When the devices are made from
semiconductor heterostructures, there are many such domains
separated among themselves by interfaces where besides the step in
potential we also have to consider the discontinuity in effective
electron mass \cite{i3}. In the `60 the problem has been solved by
applying appropriate boundary conditions \cite{i4},
\begin{equation}\label{eq1}
 \psi _L =\psi _R\,, \quad
 \frac{1}{m_L }\frac{\partial \psi _L }
  {\partial x}=\frac{1}{m_R }
  \frac{\partial \psi _R }{\partial x}\,,
\end{equation}
where {$\psi $}$_{L}$, $m_{L}$ and {$\psi $}$_{R}$, $m_{R}$ are
the electron wave function and effective mass of the electron to
the left ($L$) and right ($R$) side of a given interface. The
results obtained with the above-described procedure are in
agreement with experimental data and the method is widely used in
electron transport computations through semiconductor
heterostructures \cite{i5}. The advantage of this model is that
the transmission coefficient can be calculated using the simple
and elegant method of the transfer matrix \cite{V,MM}. However,
the assumed boundary conditions (\ref{eq1}) are imposed somewhat
artificially in order to conserve the particle current without to
have a deeper physical motivation.

Another attitude is to start with the Dirac equation even though
it is clear that the relativistic effects have to be very small.
Nevertheless, the relativistic linear dependence between energy
and mass could offer some technical advantages for finding
appropriate connection conditions at interfaces where the
potential and the effective mass present discontinuities. Few
years ago, the one-dimensional Dirac equation was successfully
used for treating problems with variable mass avoiding several
difficulties of the non-relativistic theory \cite{D}.

In this paper we would like to continue this study using the {\em
normalized} plane wave solutions of the three-dimensional Dirac
equation in helicity basis. Our purpose is to derive the
relativistic version of the transfer matrix method for the motion
in a fixed direction of a Dirac electron with point-dependent
effective mass, passing through rectangular barriers of arbitrary
profile. We show that the use of the Dirac equation allows one to
impose simple connection prescriptions at interfaces. However, the
price to pay for working with variable mass is that there are many
energy scales corresponding to different mass values. For this
reason we need to rescale the experimental potential if we want to
measure the energies with respect to an unique energy scale. The
rescaled potential will be considered the appropriate relativistic
potential of our problems. We show that only in this way the
non-relativistic limit of our approach recovers the results
derived from Schr\" odinger equation with the conditions
(\ref{eq1}).

The paper is organized as follows. In the second section we
present the well-known plane wave solutions of the Dirac equation
in the helicity basis. The next section is devoted to the problem
of one-dimensional rectangular barriers of any shape allowing us
to find the relativistic transfer matrix in section four. Finally,
it is shown that in the non-relativistic limit this matrix becomes
just the desired traditional one \cite{MM}.

\section{Plane waves}

Let us consider the Minkowski space-time in a frame of coordinates
$x^{\mu}$ ($\mu,\nu,...= 0,1,2,3$) and the metric $\eta={\rm
diag}(1,-1,-1,-1)$. In natural units (with $\hbar=c=1$) the time
is $x^0=t$ while the space coordinates, $x^1=x$, $x^2=y$ and
$x^3=z$, are the components of the vector $\vec{x}$. In this
frame, the relativistic quantum motion of an electron of mass $m$
and charge $-e$, in an arbitrary external electromagnetic field
$A_{\mu}$, is governed by the Dirac equation \cite{TH},
\begin{equation}
\gamma^{\mu}(i\partial_{\mu}-e A_{\mu})\psi -m\psi=0\,,
\end{equation}
that produces the conserved current (in units of $-e$)
\begin{equation}
j^{\mu}=\overline{\psi} \gamma^{\mu}\psi\,,
\end{equation}
where $\overline{\psi}=\psi^+\gamma^0$ is the Dirac adjoint of the
spinor $\psi$. In what follows, we take the $\gamma$-matrices in
the standard representation (with diagonal $\gamma^0$) \cite{TH}.

 Here we are interested to study the quantum modes in the particular
case of a space domain $D$ where $\vec{A}(x)=0$ and $eA_0(x)=
V={\rm const.}$ for any $\vec{x}\in D$. In this domain the Dirac
equation can be analytically solved and different quantum modes
can be well-defined using complete sets of commuting operators.
Thus the plane wave solutions are eigenspinors of the complete set
of commuting operators $\{E_D, \vec{P}, W\}$ constituted by the
Dirac operator, $E_D=i\gamma^{\mu}\partial_{\mu}-\gamma^0 V$,
momentum $\vec{P}=i\nabla$, and the Pauli-Lubanski operator
$W=2\vec{P}\cdot\vec{S}$. The corresponding eigenvalues, $m$,
$\vec{k}$ and  $\lambda$, define the plane wave spinor of positive
frequency, momentum $\vec{k}$, energy
$E(\vec{k})=\sqrt{m^2+\vec{k}^2}+ V$ and helicity $\lambda$ that
reads \cite{TH,BV}
\begin{eqnarray}
\psi_{\vec{k},\lambda}(x)&=&\frac{1}{\sqrt{2m}}\left(\begin{array}{r}
\sqrt{E(\vec{k})-V+m}\,\,\xi_\lambda(\vec{k})\\
\lambda\sqrt{E(\vec{k})-V-m}\,\,\xi_\lambda(\vec{k})
\end{array}\right)\nonumber \\
&&\times \, e^{-iE(\vec{k})t+i\vec{k}\cdot\vec{x}}\,.\label{s}
\end{eqnarray}
We denoted by $\xi_\lambda(\vec{k})$ the normalized Pauli spinors
of the helicity basis that satisfy
$\vec{k}\cdot\vec{\sigma}\,\xi_\lambda(\vec{k})=\lambda
|\vec{k}|\, \xi_\lambda(\vec{k})$ and
$[\xi_\lambda(\vec{k})]^+\xi_{\lambda'}(\vec{k})=\delta_{\lambda,\lambda'}$
(where $\sigma_i$ are the Pauli matrices and $\lambda=\pm 1$). One
can verify that each solution (\ref{s}) is normalized as
$\overline{\psi}_{\vec{k},\lambda}\psi_{\vec{k},\lambda'}=\delta_{\lambda,\lambda'}$
and produces the current
\begin{equation}
j=\frac{1}{|\vec{k}|}\,\overline{\psi}_{\vec{k},\lambda}
(\vec{k}\cdot\vec{\gamma})\,\psi_{\vec{k},\lambda}=\frac{|\vec{k}|}{m}\,,
\end{equation}
along the direction $\vec{k}$.

\section{One-dimensional motion}

\begin{figure}
\includegraphics[width=8cm]{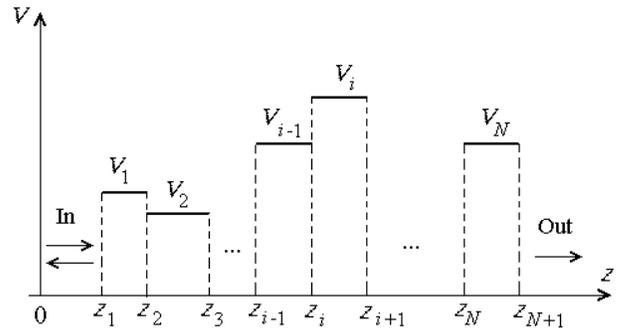}
\caption{A sequence of potential steps.}\label{fig1}
\end{figure}

The general results presented above help us to write down the
solutions of simpler one-dimensional problems along the third axis.
Of a special interest is the problem of the electron moving through
a system of $N$ rectangular barriers of arbitrary shape. In general,
the system of barriers is constituted by $N$ domains
$D_i=[z_i,z_{i+1}]$ where the potential $V(z)$ takes constant values
$V_i$ (Fig\ref{fig1}). These domains are limited by  plane
interfaces at fixed points, $z_1, z_2,..., z_{N+1}$, among them
those from $z_1$ and $z_{N+1}$ represent the interfaces between the
system of barriers and the domains outside, denoted by $D_{in}\equiv
D_0=(-\infty, z_1]$ and, respectively, $D_{out}\equiv
D_{N+1}=[z_{N+1},\infty)$. It is natural to consider that in these
latter domains the potential vanishes, $V_{in}= V_{out}=0$. In
addition, we assume that in each domain $D_i$ the electron has the
{\em effective} mass $m_i$ while in the domains $D_{in}$ and
$D_{out}$ its mass is just the {\em bare} mass $m$.

In special relativity the energy scale depends on the value of the
rest mass while the electromagnetic potential is defined up to a
gauge. Therefore, in problems where this mass is replaced by a
point-dependent effective mass, we could introduce an {\em unique}
energy scale only by choosing suitable gauge fixings, dealing with
the different values of the effective mass. In these conditions we
are encouraged to consider in each domain $D_i$ the {\em
relativistic } potential $\hat V_i$ instead of the experimental one
$V_i$. The relation among these potentials has to be derived from a
natural supplemental condition which will fixe up the gauge in the
domains $D_i$.

In any domain $D_i$ there exists a plane wave solution of energy $E$
and helicity $\lambda$ propagating in the sense of the positive
semiaxis $z$,
\begin{equation}\label{sol1}
\phi_{E,\lambda}^{i}(t,z)=\frac{1}{\sqrt{2m_i}}\left(\begin{array}{r}
k_i^{(+)}\,\xi_\lambda\\
\lambda k_i^{(-)}\,\xi_\lambda
\end{array}\right)e^{-iEt+ik_i z}\,,
\end{equation}
which depends on the constants $k_i^{(\pm)}=\sqrt{E-\hat V_i\pm
m_i}$ and scalar momentum
\begin{equation}\label{ki}
k_i={k_i^{(+)}k_i^{(-)}}=\sqrt{(E-\hat V_i)^2-m_i^2}\,.
\end{equation}
We note that in this case the helicity spinors coincide to those
of the spin basis since the spin is projected on the third axis.
Consequently, the two-component spinors $\xi_{\lambda}$ take the
usual form $\xi_1=(1,0)^T$ and $\xi_{-1}=(0,1)^T$. The plane wave
solution with the same $E$ and $\lambda$ but propagating in the
opposite sense reads
\begin{equation}\label{sol2}
\chi_{E,\lambda}^{i}(t,z)=\frac{1}{\sqrt{2m_i}}\left(\begin{array}{r}
k_i^{(+)}\,\xi_\lambda\\
-\lambda k_i^{(-)}\,\xi_\lambda
\end{array}\right)e^{-iEt-ik_i z}\,.
\end{equation}
The conclusion is that, in a domain $D_i$, the most general plane
wave solutions of energy $E$ and helicity $\lambda$ are given by
the linear combinations
\begin{equation}\label{sol}
\Psi_{E,\lambda}^{i}(t,z)=A_i \phi_{E,\lambda}^{i}(t,z) + B_i
\chi_{E,\lambda}^{i}(t,z)\,,
\end{equation}
where $A_i$ and $B_i$ are arbitrary complex numbers. Each solution
(\ref{sol}) gives the total current
\begin{equation}\label{j}
j_i=\frac{k_i}{m_i}\left(|A_i|^2-|B_i|^2\right)\,,
\end{equation}
which does not depend on helicity.

Finally we can establish the relation among the relativistic and
experimental potentials assuming that in a domain $D_i$ the
momentum $k_i$ vanishes only when the total non-relativistic
energy $E_{nr}=E-m$, calculated with respect to the bare mass $m$,
equals the experimental potential $V_i$. Therefore, according to
Eq. (\ref{ki}) we obtain the form of our relativistic potentials
\begin{equation}\label{VR}
\hat V_i=V_i +\delta m_i\,,
\end{equation}
where $\delta m_i=m-m_i$.

\section{The transfer matrix}

In what follows we shall derive the transfer matrix in the pure
scattering case without wells or tunneling  effects. This means that
the energy satisfies the condition $E \ge E_0={\rm sup}\{V_i+m
\,|\,{i=0,1,...,N+1}\}$ and $k_i$ take only real values. In
addition, we specify that the global solutions we consider here have
the same fixed energy $E$ and helicity $\lambda$ in all the domains
$D_i$. This correspond to the experimental situation when the
interactions able to produce spin-flip are absent.

In problems involving many domains $D_i$ it is difficult to
manipulate solutions of the form (\ref{sol}). For this reason we
replace these solution by {\em associated} two-dimensional vectors
\cite{V,D},
\begin{equation}\label{v}
v_i(z)=\left(\begin{array}{l} A_i e^{ik_i z}\\
B_ie^{-ik_i z}
\end{array}\right)\,,
\end{equation}
which carry all the information we need for calculating the
currents (\ref{j}). Indeed, we observe that these currents can be
expressed only in terms of $v_i(z)$ as
\begin{equation}\label{jv}
j_i=\frac{k_i}{m_i}\,[v_i(z)]^+\sigma_3v_i(z)\,,\quad \forall \,
z\in D_i\,.
\end{equation}
Thus the vectors (\ref{v}) become the basic elements of the
relativistic formalism of the transfer matrix for rectangular
barriers \cite{MM}. In the domains $D_{0}$ and $D_{N+1}$, where the
potential vanishes and the mass is $m$, the spinors
$\Psi^{0}_{E,\lambda}$ and $\Psi^{N+1}_{E,\lambda}$ have the general
form given by Eqs. (\ref{sol1}), (\ref{sol2}) and (\ref{sol}) with
the same momentum, $k_{0}=k_{N+1}=k= \sqrt{E^2-m^2}$, and constants,
$k_{0}^{(\pm)}=k_{N+1}^{(\pm)}=\sqrt{E\pm m}$. These spinors are
associated to the vectors
\begin{eqnarray}
v_{in}(z)&\equiv& v_0(z)=\left(\begin{array}{l} A_{in} e^{ik z}\\
B_{in}e^{-ik z}
\end{array}\right)\,,\\
v_{out}(z)&\equiv& v_{N+1}(z)=\left(\begin{array}{l} A_{out} e^{ik z}\\
B_{out}e^{-ik z}
\end{array}\right) \,.
\end{eqnarray}
 Now, the problem is to find the transfer matrix, $M$,  which
 transforms the $out$ vector into the $in$ one as
\begin{equation}
v_{0}(z_1)= M \, v_{N+1}(z_{N+1})\,,
\end{equation}
allowing one to calculate the transmission coefficient.

The global solution of energy $E$ and helicity $\lambda$  is
continuous in each point $z_i$ which means that we must impose the
conditions
\begin{equation}
\Psi_{E,\lambda}^{i-1}(t,z_i)=\Psi_{E,\lambda}^{i}(t,z_i)
\end{equation}
for  $i=1,2,...,N+1$. After a few manipulation we find that these
conditions lead to simple relations among the associated vectors,
\begin{equation}\label{vMv}
v_{i-1}(z_i)=M_i v_i(z_i)\,,\quad i=1,2,...,N+1\,,
\end{equation}
where the matrices
\begin{equation}
M_i=\frac{1}{2}\left(\begin{array}{cc}
r^{(+)}_i+r^{(-)}_i&r^{(+)}_i-r^{(-)}_i\\
r^{(+)}_i-r^{(-)}_i&r^{(+)}_i+r^{(-)}_i \end{array}\right)
\end{equation}
depend on the constants
\begin{equation}
r^{(+)}_i=\sqrt{\frac{m_{i-1}}{m_i}}\,\frac{k^{(+)}_i}{k^{(+)}_{i-1}}\,,\quad
r^{(-)}_i=\sqrt{\frac{m_{i-1}}{m_i}}\,\frac{k^{(-)}_i}{k^{(-)}_{i-1}}\,.
\end{equation}
 The last step is to introduce the translation matrices
\begin{equation}
T_i=\left(\begin{array}{cc}
e^{-ik_i(z_{i+1}-z_i)}&0\\
0&e^{ik_i(z_{i+1}-z_i)}
 \end{array}\right)
\end{equation}
which transform $v_i(z_i)$ into $v_i(z_{i})=T_i v_i(z_{i+1})$. With
these elements we can write down the final expression of the {\em
relativistic} transition matrix
\begin{equation}
M=\left[\prod_{i=1}^{N}M_i T_i\right]M_{N+1}\,.
\end{equation}

Let us observe that for $E\ge E_0$, when $k_i$ are real numbers,
the matrices $M_i=M_i^+$ defined by Eq. (\ref{vMv}) have the
property
\begin{equation}
M_i\sigma_3 M_i=r_i^{(+)}r_i^{(-)}\sigma_3=\frac{m_{i-1}}{k_{i-1}}
\frac{k_i}{m_i}\,\sigma_3
\end{equation}
which guarantees the conservation of the total current,
$j_{in}=j_1=...=j_i=...=j_{out}$, calculated according to Eq.
(\ref{jv}). In these circumstances, taking $B_{out}=0$ we have
$|A_{in}|^2-|B_{in}|^2=|A_{out}|^2$ which allows us to define the
transmission coefficient
\begin{equation}
{\cal T}=\frac{|A_{out}|^2}{|A_{in}|^2}=|M_{11}|^{-2}\,.
\end{equation}
We note that ${\cal T}$ results to be a function only of energy,
being independent on the helicity of the electron passing through
the rectangular barriers.

Of course, the function ${\cal T}(E)$ calculated here is defined
only on the domain $E \ge E_0$. However, starting with the present
theory, the extension to energies smaller than $E_0$ can be done but
this requires specific treatment because of the wells producing
discrete energy levels or tunneling effects which need to be treated
with specific methods.

\section{Conclusions}

Here we constructed the relativistic version of the transfer
matrix for the Dirac electron moving through  rectangular
barriers, in a similar manner as in the non-relativistic theory
based on the Schr\" odinger equation . Our approach allows one to
calculate the transfer matrices using the same rules but with
matrices $M_i$ of different forms.

Let us see what happens with our theory in the non-relativistic
limit, for small values $E_{nr}-V_i\ll m_i$. In this limit the
quantities $k^{(-)}_i = \sqrt{E_{nr}-V_i}$ remain unchanged but we
have $k^{(+)}_i \to \sqrt{2m_i}$. Consequently, we find that
$r_i^{(+)}\to 1$ remaining with the terms
\begin{equation}
r_i^{(-)}=
r_i=\sqrt{\frac{m_{i-1}}{m_i}\frac{E_{nr}-V_i}{E_{nr}-V_{i-1}}}\,,
\end{equation}
which {\em coincide} to those of Refs. \cite{MM} at least in the
domain $E\ge E_0$ considered here. Thus, the general conclusion is
that the non-relativistic limit of our approach based on the
three-dimensional Dirac equation with the relativistic potentials
(\ref{VR}) reproduces identically the results of the traditional
method based on the Schr\" odinger equation and conditions
(\ref{eq1}). Moreover, relativistic corrections can be also
calculated but so far these seem to be small in the usual regime
the electronic devices work.

In other respects, the results obtained here indicate that the use
of the Dirac equation could be helpful in other problems
concerning the motion of electrons in semiconductor
heterostructures as suggested in Ref. \cite{D}.

\subsection*{Acknowledgments}

We should like to thank Erhardt Papp for helpful discussions and
suggestions helping us to formulate our conclusions.  This work is
partially supported by CEEX - 6105 Program, Romania.

\end{document}